\documentclass[reprint,prl,showpacs,floatfix]{revtex4-1}
\usepackage{graphicx}

\begin{document}

\title{Three-Dimensional Light Bullets in Arrays of Waveguides}

\author{S. Minardi$^1$, F. Eilenberger$^1$, Y. V. Kartashov$^2$, A. Szameit$^3$, U. R\"opke$^4$, J. Kobelke$^4$\\ K. Schuster$^4$, H. Bartelt$^4$, S. Nolte$^1$, L. Torner$^2$, F. Lederer$^5$, A. T\"unnermann$^1$, and T. Pertsch$^1$}

\affiliation{$^1$ Institute of Applied Physics, Friedrich-Schiller-Universit\"at Jena, Max-Wien-Platz 1, 07743 Jena, Germany}
\affiliation{$^2$ ICFO - Institut de Ciences Fotoniques, and Universitat Politecnica de Catalunya, Mediterrean Technology Park, 08860 Castelldefels (Barcelona), Spain}
\affiliation{$^3$ Solid State Institute and Physics Department, Technion, 32000 Haifa, Israel.}
\affiliation{$^4$ Institute of Photonic Technology, Albert-Einstein-Strasse 9, 07745 Jena, Germany}
\affiliation{$^5$ Institute of Condensed Matter Theory and Solid State Optics, Friedrich-Schiller-Universit\"at Jena, Max-Wien-Platz 1, 07743 Jena, Germany}

\begin{abstract}
We report the first experimental observation of 3D-LBs, excited by femtosecond pulses in a system featuring quasi-instantaneous cubic nonlinearity and a periodic, transversally-modulated refractive index. Stringent evidence of the excitation of LBs is based on time-gated images and spectra which perfectly match our numerical simulations. Furthermore, we reveal a novel evolution mechanism forcing the LBs to follow varying dispersion/diffraction conditions, until they leave their existence range and decay.
\end{abstract}


\maketitle

\twocolumngrid

Since their theoretical prediction\cite{Silberberg90}, light bullets (LB) constitute a frontier in nonlinear science. These solitary waves are spatiotemporally localized and, in particular, find their manifestation in three-dimensional (3D) systems governed by the Nonlinear Schr\"odinger Equation (NLSE), such as quantum gases \cite{BEC}, evaporating black-holes \cite{fiberHawking}, and solid state physics \cite{Dauxois06}. 
In contrast to the one-dimensional case (e.g. optical fibers \cite{KivsharAgrawal}), where the NLSE is fully integrable and supports soliton solutions, higher dimensional solitary waves and especially LBs are not solitons in the strict sense of integrability of the dynamic equations, thus being subject to instability \cite{RasmussenBerge} and limited exsitence range \cite{Mihalache04}.
Their appeal as particle-like wavepackets triggered a two-decades-long research for a stabilization mechanism enabling full-dimensional, nonlinear light localization. Theory shows that LBs can be stabilized by a variety of experimentally motivated modifications of the NLSE, such as saturation of the nonlinearity \cite{Blagoeva91}, higher order diffraction/dispersion \cite{AkhmedievFibich}, or nonlocal nonlinearity \cite{Turitsyn85}. Transversally modulated, nonlinear media \cite{LedererReview}, \textit{e.g.} arrays of evanescently coupled waveguides, have also been  predicted to support stable LBs \cite{Aceves94,Mihalache04}. 
Despite theoretical progress, LBs eluded experimental observation. Experiments designed to observe spatiotemporal localization in nonlinear planar media \cite{EisenbergMorandotti} revealed spatiotemporal compression, but the complexity of the observations could not be readily associated to LBs. 
The complexity is due to existing optical media which support LBs only for conditions where effects beyond the Kerr-nonlinearity are influential, \textit{e.g.} self-steepening and intrapulse Raman scattering. 
For this reasons, spatio-temporal solitary waves were first observed in $\chi^{(2)}$ media where higher order effects were made negligible by artificially enhancing temporal dispersion by means of the tilted pulse technique \cite{WiseDiTrapani02}.
We remark that even in far-from-ideal systems the concept of solitary waves is a useful tool, allowing the understanding of complex nonlinear phenomena such as optical rogue waves \cite{Solli}, and super-continuum generation \cite{UME, Skryabin}. In fact, the name 'quasi-soliton' was recently used to describe nearly stationary spatiotemporal wave-packets propagating  in planar 
waveguides arrays \cite{Gorbach}.

In this Letter, we report the first observation of 3D-LBs in a two-dimensional array of coupled waveguides. 
We have found that due to higher order effects, LBs evolve following varying dispersion/diffraction conditions until they leave their existence range \cite{Mihalache04} and decay. 

\begin{figure}
\includegraphics[width=8.3 cm]{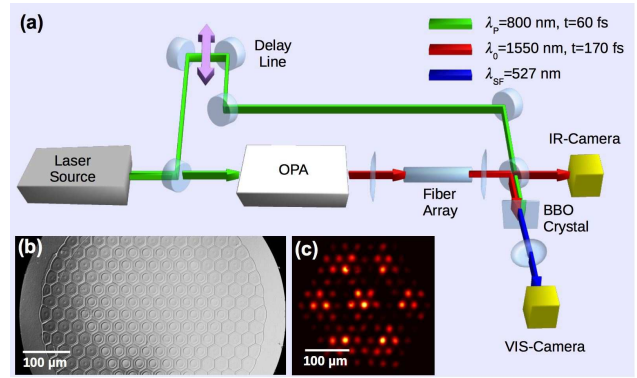}
\caption{\label{setup} (a) Layout of the experiment. A focused 170 fs infrared pulse centered at $\lambda_0=1550$ nm excites the central waveguide of a sample of the waveguide array (b). The output radiation of the sample is characterized by an infrared camera (IR-Camera) and by a spatially resolved optical gating \textit{via} frequency mixing in a 25-$\mu$m-thin BBO crystal with a delayed 60 fs probe pulse at $\lambda_p=800$ nm. (c) Experimental discrete linear diffraction pattern at the end of a 50 mm-long sample.
Parameters of the sample: array made of 91 silica cores embedded in a Fluoride-doped silica glass; lattice period of $d = 33.2 \pm 0.4 \mu$m, a core radius of $r = 9.8 \pm 0.2  \mu$m, and an index contrast of $\Delta n = 1.2\cdot10^{-3}$.}
\end{figure}

\begin{figure*}

\includegraphics[width=17cm]{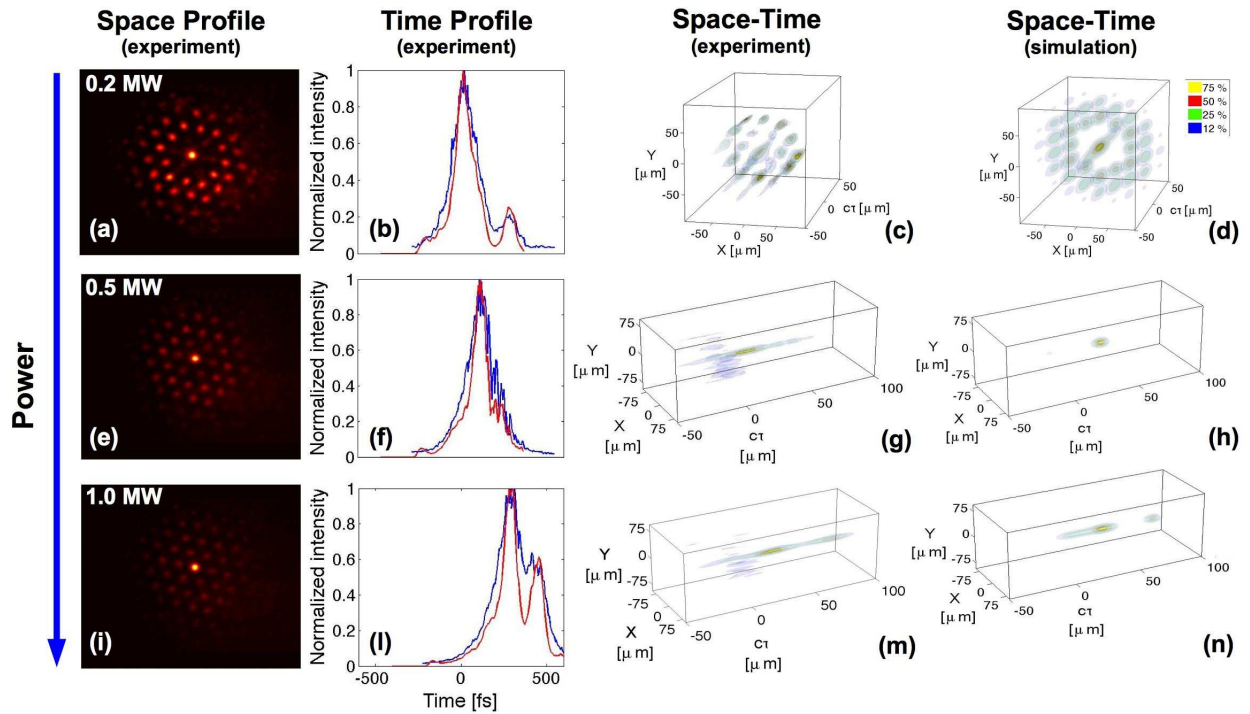}

\caption{\label{results} Experimental observation and simulation of LBs for a 40-mm-long waveguide array. Spatial infrared camera image (first column). Temporal cross-correlation trace of the central waveguide (second column) with the original experimental data (blue line) and after  deconvolution of the probe pulse profile (red line). Experimental spatiotemporal images (third column) and corresponding simulated spatiotemporal images derived (fourth column). Surfaces in the spatiotemporal images are normalized isointensity levels as indicated in the legend in (d).The three rows correspond to the three different input peak powers. 
}
\end{figure*}

The experimentally investigated system consists of an hexagonal array of evanescently coupled single mode fibres, where a focused femtosecond pulse is used to excite the central waveguide.  A microscopic image of the array cross-section is shown in Fig. 1 along with a sketch of the experimental setup. The high regularity and symmetry of the observed diffraction pattern (Fig. 1(c)) proves the unprecedented quality of the array \cite{Roepke}, and is a key for the observation of LBs, which was possible only with a high-resolution, spatially-resolved optical pulse cross-correlation techniques \cite{Potenza04,Trebino} (see Fig. 1(a)).

Temporal dispersion is anomalous around the excitation wavelength $\lambda_0=1550\:\mathrm{nm}$, but terms beyond the parabolic approximation cannot be neglected. Spatial discrete diffraction originating from inter-waveguide coupling \cite{LedererReview} is also inherently wavelength dependent. 

Figure 2 summarizes the results of experiments where the central waveguide of a 40 mm-long array of fibres was excited with 170 fs pulses. 
The spatial profile, the intensity cross-correlation trace of the central waveguide, the experimental, and simulated spatiotemporal profiles of the output wavepackets are shown for three different input power levels.
At low input power ($P_0 = 0.2$ MW) light spreads into the neighbouring waveguides while the pulse profile broadens (Fig. \ref{results} a--d). By raising the input power to $P_0 = 0.4$ MW, we observe a sharp localization of light in the initially excited  waveguide with a pulse compression below the resolution of the cross-correlator ($\approx 60$ fs, see Fig. \ref{results} e--h), while visible radiation is observed even by the naked eye. For higher powers ($P_0=1.0$  MW), the light localization is enhanced and multiple short peaks are observed in the cross-correlations, suggesting the formation of bullet-like entities.

\begin{figure}
\includegraphics[width=8.3cm]{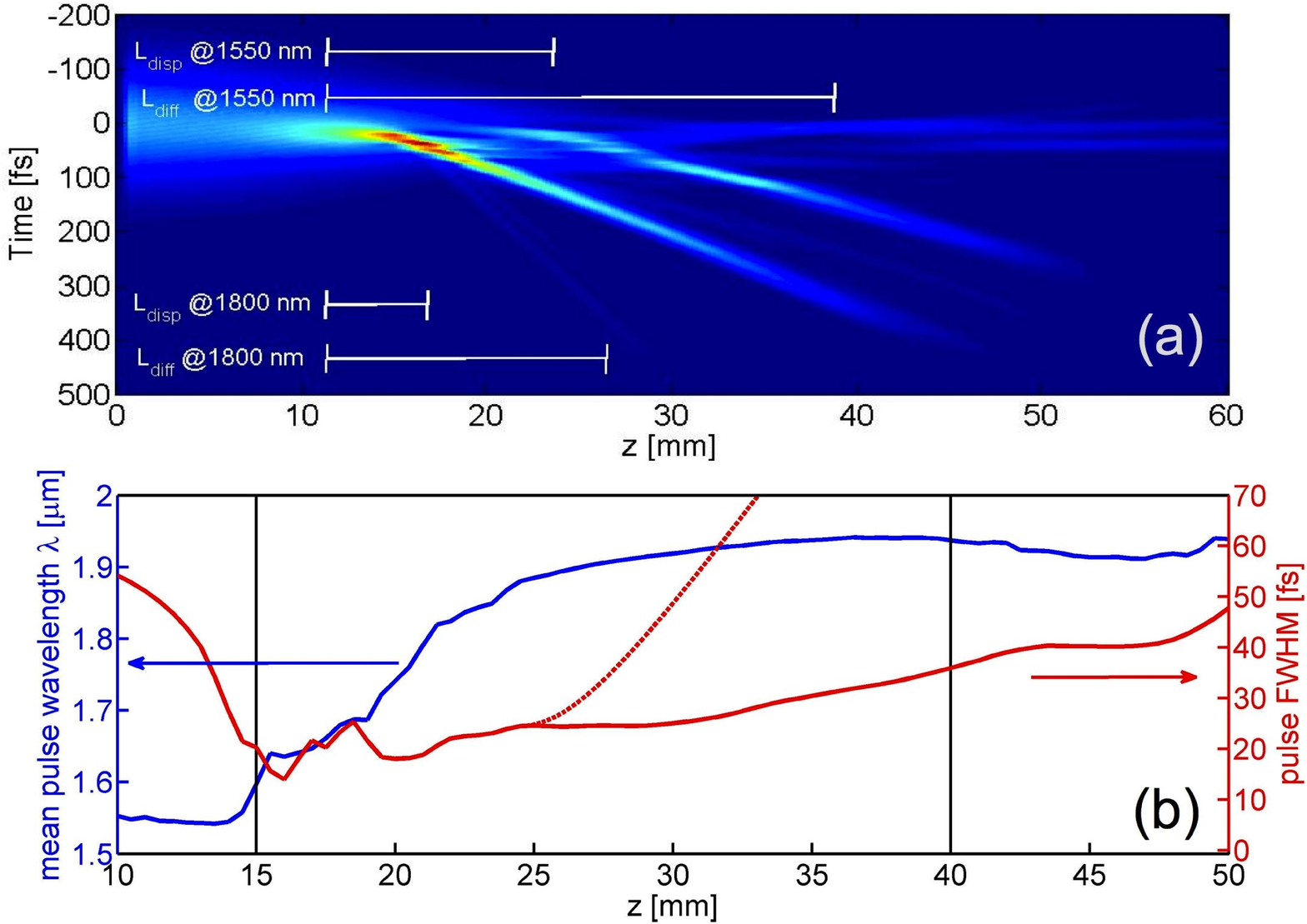}
\caption{\label{simulation} Numerical propagation of LBs in the central waveguide for $P_0 = 0.9$ MW. (a) Wavepacket evolution in the time-propagation plane. The diffraction and dispersion length corresponding to a $\tau_{\mathrm{FWHM}}=25 $ fs pulse for the minimal and maximal central wavelength of the LB are overlaid to the graph for comparison. (b) Plot of pulse duration and central wavelength for the brightest LB seen in (a).}
\end{figure} 

Numerical simulations are based on the Unidirectional Maxwell Equations \cite{UME}, which describe accurately the propagation of light in the array. Dispersion to all orders, wavelength-dependent discrete diffraction, Kerr nonlinearity, intrapulse Raman scattering and self-steepening are accounted for. The remarkable agreement with the experiment (compare Fig. 2 (d), (h), and (n) with Fig. 2 (c), (g), and (m)) did not require the adjustment of free parameters, thus justifying the use of numerical simulations to show that the observed localized wavepackets are indeed LBs.
Fig. 3(a) displays the evolution of the temporal profile of the light in the central waveguide of the array for an input power of $P_0=0.9$ MW. Within the first 15 mm of propagation, the excited pulse gets spatiotemporally focused and splits into several 
fragments with (FWHM) durations ranging from 15 to 30 fs.
The central wavelength of the fragments increases rapidly to larger values in proportion to their initial intensity, thus accounting for the observed distribution of group velocities. The brightest pulse experiences several collisions and eventually 
propagates as virtually stationary LBs to a distance of $z=40\:\mathrm{mm}$ (see Fig. 3(b)), beyond which the wavepacket suddenly spreads in space and time. Because the central wavelength of the LB is not constant and the characteristic dispersion and diffraction lengths ($L_{\mathrm{disp}}$, $L_{\mathrm{diff}}$) are wavelength dependent, we can gauge the roboustness of the localization of light in the array by defining an average dispersion and diffraction length $\langle L_{\mathrm{disp}} \rangle$ and $\langle L_{\mathrm{diff}} \rangle$ for the brightest pulse  (see definition at the end of the Letter). According to this definition, nearly stationary propagation is achieved over 1.9 $\langle L_{\mathrm{diff}} \rangle$ and 9.0 $\langle L_{\mathrm{disp}} \rangle$.
We also proved that the light localization is not due to conical waves \cite{PaoloX} by numerically verifying that the wavepackets do not self-heal.

This scenario of the LB evolution has been confirmed experimentally by means of Frequency Resolved Optical Gating (FROG) \cite{Trebino} obtained from a set of spatiotemporal images, each recorded by inserting after the sum-frequency crystal a 10-nm interferential filter chosen form a set of 6, with central transmission wavelength ranging from 510 to 560 nm. This is equivalent to sampling the time-frequency plane of the LBs between $\lambda=1423$ nm and $\lambda=1870$ nm. In the left column of Fig. 4, we show the experimental FROG traces for the pulse in the central waveguide of 
three samples of the array (length $L = 25$, 40 and 60 mm, $P_0=0.9$ MW). Frequency-wise interpolation was used to display the data. Two LBs with central wavelengths of 1650 nm and 1810 nm are observed at $z = 25$ mm. Only a LB with $\lambda_c = 1790$ nm is observed at $z = 40$ mm. 
The experimental data can be interpreted by the corresponding simulated FROG traces (Fig. 4, right column), which accurately reproduce the observed features. The simulation reveals that the long-wavelength LB observed at $z = 25$ mm is red-shifted to $\lambda_c = 1950$ nm (i.e. beyond the spectral acceptance of our FROG) while the bluest is actually the one observed at z = 40 mm.
Further propagation to $z = 60$ mm leads to full decay of the LBs captured in the measurements at $z = 25$ and $z = 40$ mm, and only weaker, moderately red-shifted fragments surviving collisions in the central wavepacket are observed.

\begin{figure}
\includegraphics[width=8.3cm]{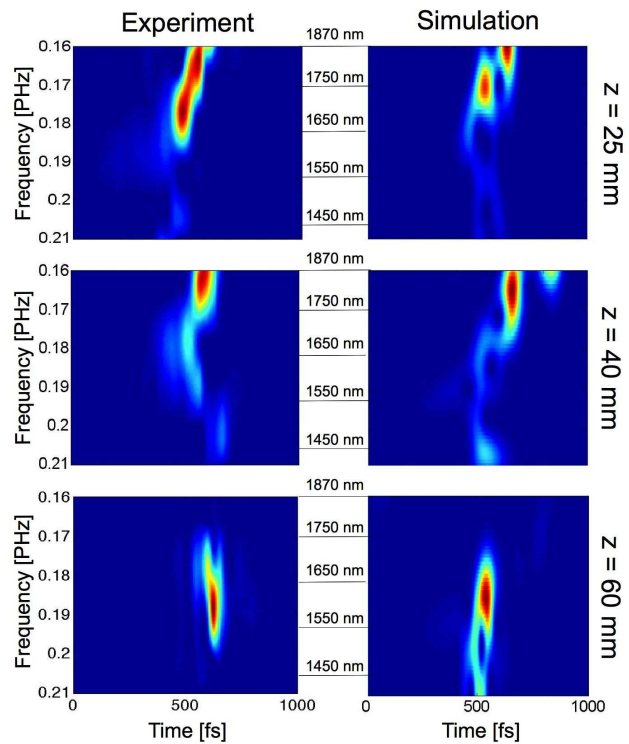}
\caption{\label{simulation} Experimental (left) and simulated (right) FROG traces of  the pulse propagating in the central waveguide measured at the output of samples. Top: $L = 25$ mm. Center: $L = 40$ mm. Bottom: $L =  60$ mm. The wavelength scale is shown between the columns. Input power $P_0 = 0.9$ MW.}
\end{figure}  

\begin{figure}
\includegraphics[width=8.3cm]{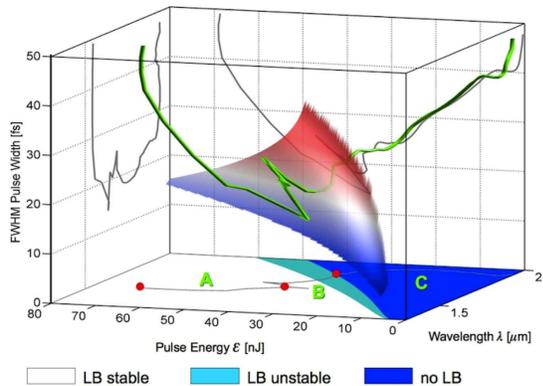}
\caption{\label{theory} Green line: trajectory in parameter space of the brigthest LB appearing in the simulation of Fig. 3a. The blue-red surface shows the locus of the ideal stationary LB solution of the discrete NLSE. Energy/wavelength plane: regions of stability and existence of ideal LB; A-C indicate different phases of the trajectory (see text for details).  Gray lines: projection of the trajectory onto the orthogonal planes.}
\end{figure}

The key to understand the LBs decay mechanism is to consider the influence of the wavelength-dependent dispersion and diffraction parameters of our system. While propagating through the sample, the LB is red-shifted due to self-steepening and Raman scattering. In this course, the LB reshapes adiabatically to the increased dispersion and diffraction strength pertaining to the shifted wavelength. Because for longer wavelengths the coupling strength between waveguides increases exponentially, the energy threshold \cite{Mihalache04} required for the LB propagation grows steeply. Therefore the solitary wave will eventually decay when the dispersively modified energy threshold exceeds the energy of the excitation. The scenario is illustrated in Fig. 5. The curved surface defines the temporal width of the idealized 
LB solution of the discrete NLSE \cite{Aceves94} without self-steepening and Raman scattering, determined uniquely for each carrier wavelength $\lambda$ and energy $E$ of the central waveguide component of the LB. Additionally, the trajectory of the brightest pulse appearing in the dynamic simulation in Fig. 3(a) is overlaid to the plot (green line) along with its orthogonal projections (gray lines). After strong pulse reshaping (region A), the trajectory relaxes to the surface of the ideal LBs (region B), until the redshift pulls it behind the surface, where no LB can exist (region C). Because of the finite energy threshold for the existence of 3D discrete-continuous LBs \cite{Mihalache04} this decay scenario is quite generic for experimental systems. Notice that the trajectory of the bullet in the $E-\lambda$ plane indicates that radiation losses are rather weak. To show that the decay mechanism is independent from radiation losses, we performed a simulation in an almost ideal Kerr medium with pure quadratic chromatic dispersion, and longer coupling length ($L_{\mathrm{diff}}$(1550 nm)=96 mm). These artificial conditions ensure that radiation from the LBs is strongly suppressed by large phase mismatch \cite{Skryabin} and lower power excitation threshold, due to weaker coupling. The red shifting mechanism is retained by including self-steepening and Raman scattering. The results of the simulation reveal essentially the same dynamics as depicted in Fig. 5, while the propagation range of the LB normalized to $L_{\mathrm{diff}}$ is expanded only by a factor of 1.5 (3 $\langle L_{\mathrm{diff}}\rangle$ in total). 

In conclusion, we reported for the first time the experimental observation of 3D nonlinear LBs. Our work revealed a new evolution scenario of adiabatic reshaping of LBs which sheds new light on previously puzzling results obtained in similar systems \cite{EisenbergMorandotti}, and offers an interpretation in terms of evolving LBs for the dynamic of the so-called spatiotemporal 'quasi-solitons' \cite{Gorbach}.
We point out that the identification of the decay mechanism opens new perspectives for the design of systems aimed at the excitation of very long-lived LBs.

\paragraph{Definition of average diffraction and dispersion length.}
The average dispersion length over the path from $z_1$ to $z_2$ is defined as:
\begin{equation}
\langle L_{\mathrm{disp}} \rangle=\frac{1}{z_2-z_1} \int_{z_1}^{z_2}\frac{\tau_0(\tilde{z})^2}
{\parallel \left[{d^2\beta}/{d\omega^2}\right]_{\omega_{c}(\tilde{z})} \parallel}
d\tilde{z}
\end{equation}
where $\beta(\omega)$ is the propagation constant of light in the single waveguide, $\tau_0(\tilde{z})$ and $\omega_c(\tilde{z})$ are the $1/e$ pulse duration and the carrier angular frequency of the bullet at the propagation distance $\tilde{z}$, respectively.
Analogously, we define the average diffraction length on the same path as:
\begin{equation}
\langle L_{\mathrm{diff}}\rangle=\frac{1}{z_2-z_1}\int_{z_1}^{z_2}\frac{\pi}{2\sqrt{6}c(\omega_c(\tilde{z}))}d\tilde{z} 
\end{equation}
which relates to the waveguide coupling strength $c(\omega)$ \cite{LedererReview}. For the simulation of Fig. 3(a): $\langle L_{\mathrm{disp}}\rangle=2.6$ mm, $\langle L_{\mathrm{diff}}\rangle=12.7$ mm.


\begin{thebibliography}{999}
\bibitem{Silberberg90} Silberberg, Y. \emph{Opt. Lett.} \textbf{22}, 1282--1284 (1990).
\bibitem{BEC} Strecker, K. E. \emph{et al.}, \emph{Nature} \textbf{417}, 150--153 (2002).
\bibitem{fiberHawking} Philbin, T. G.  \emph{et al.},
\emph{Science} \textbf{319}, 1367--1370 (2008).
\bibitem{Dauxois06} Dauxois, T., Peyard, M. \emph{Physics of Solitons}. Cambridge, Cambridge (2006).
\bibitem{KivsharAgrawal} Kivshar, Y. S., Agrawal, G. \emph{Optical Solitons: From Fibers to Photonic Crystals}. Academic Press, London (2006).
\bibitem{RasmussenBerge} Rasmussen, J. J., Rypdal, K. 
\bibitem{Mihalache04} Mihalache, D., \emph{et al.}
\emph{Phys. Rev. E} \textbf{70}, 055603 (2004).
\emph{Physica Scripta} \textbf{33}, 481 
(1986); Berge, L. \emph{Physics Reports} \textbf{303}, 259
(1998).
\bibitem{Blagoeva91} Blagoeva, A.B., \emph{et al.} \emph{IEEE Journal of Quantum Electronics} \textbf{27}, 2060--2065 (1991).
\bibitem{AkhmedievFibich} Akhmediev, N., \emph{et al.}, \emph{Opt. Lett.} \textbf{18}, 411--413 (1993); Fibich, G., Ilan, B. \emph{Opt. Lett.} \textbf{29}, 887--889 (2004).
\bibitem{Turitsyn85} Turitsyn, S. K. \emph{Theor. Mat. Fiz.}, \textbf{64}, 226--232 (1985).
\bibitem{LedererReview} Lederer, F., \emph{et al.} 
\emph{Phys. Rep.} \textbf{463}, 1--126 (2008).
\bibitem{Aceves94} Aceves, A. B., \emph{et al.}
\emph{Opt. Lett.} \textbf{19}, 329--331 (1994).
\bibitem{EisenbergMorandotti} Eisenberg, H. S., \emph{et al.} 
\emph{Phys. Rev. Lett.} \textbf{87}, 043902 (2001); Cheskis, D., \emph{et al.} \emph{Phys. Rev. Lett.} \textbf{91}, 223901 (2003).
\bibitem{WiseDiTrapani02} Wise, F., and Di Trapani, P. The hunt for light bullets - Spatiotemporal solitons. \emph{OPN}, \textbf{28}, Feb. (2002).
\bibitem{Solli} Solli, D. R., \emph{et al.}  \emph{Nature} \textbf{450}, 1054--1058 (2007).
\bibitem{UME} Husakou, A. V., Herrmann, J. . 
\emph{Phys. Rev. Lett.} \textbf{87}, 203901 (2001).
\bibitem{Skryabin} Yulin, A. V.,\emph{et al.} \emph{Opt. Lett.} \textbf{29}, 2411--2413 (2004).
\bibitem{Gorbach} Benton, C. J., Gorbach, A. V., Skryabin, D. V. \emph{Phys. Rev. A} \textbf{78}, 033818 (2008).
\bibitem{Roepke} R\"opke, U., \emph{et al.}
\emph{Opt. Exp.} \textbf{15}, 6894--6899 (2007). 
\bibitem{Potenza04} S. Minardi , J. Trull, M.A.C. Potenza J. of Holography and Speckles 6, 53 (2009).
\bibitem{Trebino} Trebino, R., Kane, D. J. \emph{J. Opt. Soc. Am. A} \textbf{10}, 1101--1110  (1993). 
\bibitem{PaoloX} Porras, M. A., \emph{et al.}\emph{J. Opt. Soc. Am. B} \textbf{22}, 1406--1413 (2005).

\end{thebibliography}
\end{document}